\documentstyle[12pt]{article}
\newcommand{\bea}{\begin{eqnarray}}
\newcommand{\eea}{\end{eqnarray}}
\newcommand{\be}{\begin{equation}}
\newcommand{\ee}{\end{equation}}
\newcommand{\vs}[1]{\vspace{#1 mm}}

\newcommand{\dsl}{\pa \kern-0.5em /}

\newcommand{\pa}{\partial}

\newcommand{\nn}{\nonumber\\}

\begin{document}
\topmargin 0pt
\oddsidemargin 0mm

%\renewcommand{\thefootnote}{\fnsymbol{footnote}}

%\begin{titlepage}

\begin{flushright}

USTC-ICTS-07-11\\

%hep-th/yymmnnn\\

%SINP-TNP/02-7

\end{flushright}

\vspace{2mm}

\begin{center}

{\Large \bf On Brane-Antibrane Forces

}

\vs{6}

{\large J. X. Lu$^a$\footnote{E-mail: jxlu@ustc.edu.cn}, Bo
Ning$^a$\footnote{E-mail: nwaves@mail.ustc.edu.cn}, Shibaji
Roy$^b$\footnote{E-mail: shibaji.roy@saha.ac.in}
 and Shan-Shan Xu$^a$ \footnote{E-mail: xuss@mail.ustc.edu.cn}}

 \vspace{4mm}

{\em

 $^a$ Interdisciplinary Center for Theoretical Study\\

 University of Science and Technology of China, Hefei, Anhui
 230026, China\\

%Michigan Center for Theoretical Physics\\

%Randall Laboratory, Department of Physics\\

%University of Michigan, Ann Arbor, MI 48109-1120, USA\\

\vs{4}

 $^b$ Saha Institute of Nuclear Physics,
 1/AF Bidhannagar, Calcutta-700 064, India}

\end{center}

\vs{15}

\begin{abstract}

In this note, we will discuss two aspects of brane-antibrane forces.
In one aspect, we generalize the force calculation of D0-${\bar {\rm D}}$0
of Banks and Susskind to D$p$-${\bar {\rm D}}p$ for $1\le p \leq
8$. In particular, we find that the force is also divergent for $p =
1$ while for the other cases ($p \ge 2$) the forces are in general
finite when $Z \to 0^+$, where $Z = \frac{Y^2}{2\pi^2\alpha'} - 1$
with $Y$, the brane-antibrane separation. However, the forces are
divergent for all cases when $Z < 0$, signalling the occurrence of
open string tachyon
condensation in this regime. The other deals with the puzzling static
nature of the supergravity brane-antibrane configurations. We will
show that the force on a brane probe due to a brane-antibrane
background vanishes when the probe is placed at the location of the
coincident brane-antibranes, thereby providing a direct evidence
for the existence of general static brane-antibrane configuration in
the supergravity approximation.

\end{abstract}
\newpage
\section{Introduction}

The brane-antibrane systems in type II superstrings break all
spacetime supersymmetries. Consequently, unlike the so-called BPS
branes, their dynamics, even though more interesting, is difficult
to understand given our limited knowledge of the underlying full
theory. A coincident D-brane-antiD-brane pair (or a non-BPS D-brane)
in type II theories is unstable due to the presence of tachyonic
mode in the weakly coupled open string description \cite{Sen:1999mg},
however, it should be noted that the corresponding strongly
coupled system can be very complicated and the underlying dynamics
may be different
\cite{Sen:1997pr, Bergman:2000xf, Lu:2005jc}. As a result, these
systems decay and the decay occurs by the process known as tachyon
condensation \cite{Sen:1998sm}. The tachyon condensation is well
understood by now in the open string description using either the
string field theory approach \cite{Sen:1999nx, Gerasimov:2000zp,
Schnabl:2005gv} or the tachyon effective action approach
\cite{Kraus:2000nj} on the brane.
The closed string (or supergravity) approach on this process and
the related issues have also been
discussed in \cite{Brax:2000cf, Lu:2004dp, Bai:2005jr, Lu:2005ju}. In this
approach, we  interpret the known non-supersymmetric ten-dimensional
type II supergravity solutions \cite{Zhou:1999nm, Ivashchuk:2001ra,
Miao:2004bn, Lu:2004ms, Bai:2005jr, Bai:2006vv} as coincident
brane-antibrane systems and relate the parameters in the solutions
to the corresponding microscopic parameters such as the number of
branes, the number of antibranes and the tachyon parameter.
Using these relations, we have calculated the ADM mass
and have shown that the solution and the ADM mass capture all the
required properties and give a correct description of the tachyon
condensation \cite{Lu:2004dp, Bai:2005jr} as advocated by Sen
\cite{Sen:1998sm} on the D-$\bar {\rm D}$ system.

In this note, we will discuss two issues on the brane-antibrane
systems, one in the open string description and the other in the
supergravity (or closed string) description. One possible way to get
a signal of the occurrence of the tachyon condensation is through
calculating the brane-antibrane force at a given separation and
examine the force behavior as the separation approaches the string
scale as Banks and Susskind did in \cite{Banks:1995ch} for
D0-$\bar{\rm D}$0 system. We will generalize Banks and Susskind's
analysis of $p=0$ to $1 \le p \le 8$\footnote{To make sense of a
separation between two D$p$ branes, we need to limit $p \le 8$ since
$p = 9$ is a spacetime filling brane.} in the following section. We
will see that $p = 1$ case is similar to $p = 0$, i.e., the force
between the brane and the antibrane is divergent, while for other $p
\ge 2$ the story is different and the force is actually finite, when
$Z \to 0^+$ where $Z = \frac{Y^2}{2\pi^2\alpha'} - 1$ with $Y$ the
brane-antibrane separation. However, when $Z < 0$, the force is
always divergent (for all $p$), indicating the occurrence of the
tachyon condensation in that regime, since the divergence is
actually due to the tachyon mode of the open string connecting the
D$p$ and the $\bar{\rm D}p$. The divergence of the Born-approximated
force between the brane and the antibrane at a separation of string
scale indicates such a description breaks down but at the same time,
the appearance of such a violent force may also indicate the
occurrence of a new process and we know that this corresponds to the
tachyon condensation. So we may use the appearance of such violent
force as an evidence for the occurrence of the tachyon condensation.

The interpretation of non-supersymmetric static solutions of ten
dimensional type II supergravities as representing the
coincident brane-antibrane systems has a puzzle. One in general
expects the non-existence of such a static configuration since the
system under consideration is unstable. Therefore the static nature
of these solutions must be due to the supergravity approximation.
This static nature has an advantage in that it enables us to use the ADM
mass (calculated asymptotically) to capture the off-shell tachyon
potential as mentioned in \cite{Brax:2000cf} and discussed further
in \cite{Lu:2006rb} even though the small distance behavior of such
solutions may not be trusted in general. In other words, the
off-shell tachyon potential can be represented by a continuous
family of ADM mass which corresponds to a family of static
supergravity configurations labeled by the mass parameter. In this
sense, we can discuss the tachyon condensation semi-empirically if
the parameters in the solutions can be related to the number of
branes, the number of antibranes and the tachyon parameter as was
achieved successfully in \cite{Lu:2004dp, Bai:2005jr}. How to
understand the static nature of the non-supersymmetric supergravity
solutions has been discussed in \cite{Brax:2000cf}. It was argued
there that these solutions can be static even if the brane sources
are time dependent, in analogy with the static exterior geometry of a
pulsating spherically symmetric star, thanks to Birkhoff's theorem,
and the time-dependence could presumably be discerned to the level
of higher mass modes of closed string. This has been further
addressed in \cite{Lu:2006rb} for the case of chargeless
configurations by considering the relation between the
disappearance of conical singularity and the vanishing force between
the coincident brane-antibrane in the supergravity approximation. In
section 3, we will use a probe approach to show that when a probe reaches
the location of the coincident branes-antibranes but still
not strongly bounded to the brane-antibrane system (in other
words, the probe can still be taken as a probe), the force between
the probe and the system vanishes, therefore providing a more direct
evidence for the static nature of a general brane-antibrane system
in the supergravity approximation.

\section{The analysis of brane-antibrane force}

We consider weakly coupled type II strings in ten dimensions. These
theories admit various BPS D$p$ branes with $p$ even in IIA
theories and odd in IIB theories. The brane tension is inversely
proportional to the string coupling $g$ and as such in the
weak-coupling limit, i.e. $g \to 0$, the tension will be divergent.
So, one may naively expect that the brane may no longer be taken as
rigid and flat and neither the spacetime can be flat any more in
contrary to what we usually do in the perturbative calculations. In
the following, we will show that the naive expectation is wrong
even up to the distance of the order of string scale. For this, let us
first consider the metric of the D$p$ brane supergravity configurations
as\cite{Duff:1993ye, Duff:1994an}, \be ds^2 = \left(1 +
\frac{k_p}{r^{7 - p}}\right)^{-\frac{7 - p}{8}} dx_{\|}^2 + \left(1
+ \frac{k_p}{r^{7 - p}}\right)^{\frac{p + 1}{8}} dx_\bot^2, \ee
where $r$ is the radial distance transverse to the brane,
$x_\|$ are the directions along the branes and $x_\bot$ are
those transverse to the branes. For asymptotically-flat and
well-behaved supergravity configurations, we need to take  $p =0, 1,
\cdots 6$. The parameter $k_p$ is related to the ten dimensional
Newton constant $2 \kappa^2$, the number of D$p$ branes $N$ and the
D$p$-brane
tension $T_p$, apart from some numerical factor (which are
irrelevant to the following discussion and will be ignored), as \be
k_p \sim 2\kappa^2 N T_p \sim N g \alpha'^{(7-p)/2} ,\ee where we
have expressed $2\kappa^2$ and $T_p$ in terms of $\alpha'$ and $g$
as given, for example, in \cite{deAlwis:1996ez}. From this, it is
clear that for a large but fixed $N$ and fixed $\alpha'$, $k_p$ vanishes
as $g\to 0$  and therefore the spacetime remains flat even
for $ \alpha'^{1/2} \gg r \gg (N g)^{1/(7 - p)} \alpha'^{1/2}$. So
if we don't probe a distance much smaller than the string scale, we
are safe to take both the brane and the spacetime as flat in the
lowest order calculation.

The calculation of the interaction (amplitude) between two parallel
D$p$ branes separated by a distance $Y$ can be computed (for example
as given in \cite{Polchinski:1995mt}) in the lowest order as an open
string one-loop annulus diagram with one end of the open string
located at one D$p$ brane and the other end at the other D$p$ brane.
This can also be viewed as a tree-level closed string amplitude,
creating a closed string at one D$p$ brane, propagating a distance
$Y$ and then being absorbed by the other D-brane at the other end.
The interaction amplitude has two contributions, one from the NSNS
closed string exchange and the other from the RR closed string
exchange. It is \bea A &=& V_{p + 1} \int_0^\infty \frac{dt}{t}
\left(2\pi t\right)^{- \frac{(p + 1)}{2}} e^{- \frac{t Y^2}{8 \pi^2
\alpha'^2}} \prod_{n = 1}^\infty \left(1 - q^{2 n}\right)^{- 8}\nn
&\,\,&\qquad\frac{1}{2}\left\{ -16 \prod_{n = 1}^\infty \left(1 +
q^{2n}\right)^8 + q^{-1} \prod_{n = 1}^\infty \left(1 + q^{2n -
1}\right)^8 - q^{-1} \prod_{n = 1}^\infty \left(1 - q^{2n -
1}\right)^8 \right\},\eea where $V_{p + 1}$ is the $p$-brane
worldvolume, $q = e^{- t/4\alpha'}$ and the integration variable $t$
is the proper time in the open string channel. In the above, the
first two terms in the curly bracket are from the NSNS closed string
sector exchange while the the third term is from the RR sector. The
BPS nature of this interaction tells that the amplitude actually
vanishes which can also be seen from the above two NSNS terms
canceling the third RR term using the ``usual abstruse identity".
The interaction for a D$p$ brane and an anti D$p$-brane placed
parallel at a separation $Y$ can be obtained from the above simply
by switching the sign in front of the RR term and  the amplitude is
therefore just twice the absolute value of the RR term and is given
as \be {\cal A} \equiv \frac{A}{ V_{p + 1}} =  \int_0^\infty
\frac{dt}{t} \left(2\pi t\right)^{- \frac{(p + 1)}{2}} e^{-
\frac{t}{4\alpha'}( \frac{Y^2}{2\pi^2\alpha'} - 1) } \prod_{n =
1}^\infty \left(\frac{1 - q^{2n - 1}}{1 - q^{2 n}}\right)^8, \ee
where we have defined ${\cal A}$, the interaction amplitude per unit
$p$-brane volume and as in \cite{Banks:1995ch}, we introduce the
parameter $Z$ as \be Z = \frac{Y^2}{2\pi^2\alpha'} - 1\ee and the
function \be g (t) = \prod_{n = 1}^\infty \left(\frac{1 - q^{2n -
1}}{1 - q^{2 n}}\right)^8.\ee One can show using the relations for
$\theta$-functions that $g (t) \to 1$ as $t \to \infty$ while $g (t)
\to (t/2\pi\alpha')^4$ as $t \to 0$. These limits will be needed
later. For simplicity, let us define a variable $u = t/4\alpha'$ and
the attractive force per unit $p$-brane volume is now \be f = -
\frac{d {\cal A}}{d Y} = \frac{Y} {\pi^2 \alpha' (8\pi\alpha')^{(p +
1)/2}} \int_0^\infty du u^{- \frac{p + 1}{2}} e^{- u Z} g (u),\ee
where \be
g (u) = \left\{\begin{array}{cc} 1&\qquad\qquad u \to \infty\\
(2 u/\pi)^4&\qquad \qquad u\to 0\end{array}\right . \ee and $0 <
g(u) <  1$ in general. For $Z > 0$, the only possible divergence for
the above force comes from $u \to 0$ and one can show using the
limiting expression for $g(u)$ in (8) for $u \to 0$ that the
integration is actually convergent there for those allowed $ 0 \le p
\le 8$. Therefore the attractive force is finite as expected since
no new process such as tachyon condensation occurs when the
brane-antibrane separation is larger than the string scale.

We now examine the force behavior when $Z \to 0^+$. For this, let us
change the integration variable to $v = Z u $, we have now \bea f
&\sim& Z^{\frac{p - 1}{2}} \int_0^\infty dv v^{- \frac{p + 1}{2}}
e^{- v} g \left(\frac{v}{Z}\right)\nn &=& Z^{\frac{p - 1}{2}}\left[
\int_0^{a Z} dv v^{- \frac{p + 1}{2}} e^{- v} g
\left(\frac{v}{Z}\right) + \int_{a Z}^\infty dv v^{- \frac{p +
1}{2}} e^{- v} g \left(\frac{v}{Z}\right) \right]\nn & \ge &
Z^{\frac{p - 1}{2}} \int_{a Z}^\infty dv v^{- \frac{p + 1}{2}} e^{-
v} g \left(\frac{v}{Z}\right)\nn &\approx& Z^{\frac{p - 1}{2}}
\int_{a Z}^\infty dv v^{- \frac{p + 1}{2}} e^{- v} ,\eea where $a$
is a fixed large number ($\gg 1$) and in the last line we have used
$g(u) \to
1$ for large $u$. Let us examine the integration in the last line
above when $Z \to 0^+$. For $p = 0$, the integration is $\Gamma
(1/2)$, finite, and the force $f \ge 1/\sqrt{Z} \to \infty$ as
discussed by Banks and Susskind in \cite{Banks:1995ch}. For $p = 1$,
the force $f \ge \Gamma (0) \to \infty$ is also divergent. Therefore
this case is similar to the $p = 0$ case. For $p \ge 2$, the above
expression for the integration appears as $0 \cdot \infty$ and we
need a more careful analysis than the above.

For this, let us re-express the force as \bea f &\sim& Z^{\frac{p -
1}{2}} \int_0^\infty dv v^{- \frac{p + 1}{2}} e^{- v} g
\left(\frac{v}{Z}\right)\nn & = & Z^{\frac{p - 1}{2}}\left[\int_0^{b
Z} dv v^{- \frac{p + 1}{2}} e^{- v} g \left(\frac{v}{Z}\right) +
\int_{b Z}^{a Z} dv v^{- \frac{p + 1}{2}} e^{- v} g
\left(\frac{v}{Z}\right) + \int_{a Z} ^\infty dv v^{- \frac{p +
1}{2}} e^{- v} g \left(\frac{v}{Z}\right)\right] \nn &=& Z^{\frac{p
- 1}{2}}\left[ \int_0^{b Z} dv v^{- \frac{p + 1}{2}} e^{- v} g
\left(\frac{v}{Z}\right) + \int_{a Z} ^\infty dv v^{- \frac{p +
1}{2}} e^{- v} g \left(\frac{v}{Z}\right)\right] + \int_b^a du u^{-
\frac{p + 1}{2}} e^{- u Z} g (u).\nn \eea where we have introduced
two fixed parameters $b$ and $a$ with $b\ll 1$ while $a\gg 1$. The
last term in the last line of eq.(10) corresponds to the second term
of the second line of the same equation. However, note that we have
expressed it in terms of the original integration variable $u$.
Since both $b$ and $a$ are fixed, so the last term in the last line
in (10) should be finite. Let us examine the first term in the
square bracket with the pre-factor $Z^{(p - 1)/2}$ in the last line
in (10). Since $b$ is very small, so we can approximate the function
$ g(v/Z) \sim (v/Z)^4 $ (as given in eq.(8)) in the integration.
With this, one can show \bea Z^{\frac{p - 1}{2}} \int_0^{b Z} dv
v^{- \frac{p + 1}{2}} e^{- v} g \left(\frac{v}{Z}\right) &\sim&
Z^{\frac{p - 1}{2}} \int_0^{b Z} dv v^{- \frac{p + 1}{2}} e^{- v}
(v/Z)^4\nn &\sim& b^{(9 - p)/2},\eea i.e., finite. Let us examine
the second term with the pre-factor now. For very large $a$, we have
\bea Z^{(p - 1)/2} \int_{a Z}^\infty dv v^{- \frac{p + 1}{2}} e^{-
v} g \left(\frac{v}{Z}\right) &\sim & Z^{(p - 1)/2} \int_{a
Z}^\infty dv v^{- \frac{p + 1}{2}} e^{- v} \nn & =& Z^{(p - 1)/2}
\left[ \int_1^\infty dv v^{- \frac{p + 1}{2}} e^{- v} + \int_{a Z}^1
dv v^{- \frac{p + 1}{2}} e^{- v}\right] \nn & < & a^{(1 - p)/2},\eea
therefore also finite as $Z\to 0^+$. Here we have used $g (u) \to 1$
for large $u$ in the first line above. Also it is obvious that the
first term in the square bracket in the second line above is finite
and the second term is \be \int_{a Z}^1 dv v^{- \frac{p + 1}{2}}
e^{- v} < \int_{a Z}^1 dv v^{- \frac{p + 1}{2}} \sim (a Z)^{(1 -
p)/2}.\ee So the force between the brane and the antibrane is finite
when $Z \to 0^+$ for $2 \le p \le 8$.

In summary, we have seen that the force between the brane and the
antibrane is
divergent for $p = 0, 1$ while it is finite for $2 \le p \le 8$ when $Z\to
0^+$. Further, the force is always divergent for $0 \le p \le 8$
when $Z < 0$. The above divergences are due to large $u$
contribution to the force given in (7) and can be understood by writing
the large $u$ behavior as $ \int^\infty du u^{-
\frac{p + 1}{2}} e^{- Z u} [1 + {\cal O} (e^{ - u})]$. Now it is clear
that the large $u$
integration diverges when $p = 0, 1$ while it converges for
$2\le p \le 8$ when $Z = 0$. When $Z < 0$, the exponential $e^{ - Z
u}$ in the integration dominates and dictates the large $u$
divergence for all $0\le p \le 8$.  These divergences are due to the
tachyon mode of the open string connecting the D$p$ and the $\bar{\rm
D}p$ as can be seen from the expansion of $g (u) = 1 + {\cal O} (e^{
- u})$ for large $u$ where the first term `1' corresponds to the
tachyon mode contribution. As discussed in the Introduction, the
appearance of such a violent force indicates the breakdown of the
calculation or it can be thought of as an indication of a new process,
therefore, signalling the occurrence of the tachyon condensation.

Another way to understand the connection between the force
divergence and the onset of tachyonic instability is as follows: The
force divergence implies the appearance of certain pole in the force
calculation. But this divergence occurs either at space-like
separation $Y^2 \le 2\pi^2 \alpha'$ for $p = 0, 1$ or at space-like
separation $Y^2 < 2\pi^2 \alpha'$ for all $p$  which implies that
the corresponding
pole is a tachyon i.e. we see the onset of tachyonic instability, since
only a tachyonic pole can propagate in a space-like separation.

The above discussion implies that the initiation of tachyonic
instability for brane-antibrane systems is different for $p \le
 1$ and for $p > 1$. For $p \le 1$, this occurs at a larger brane
 separation and the onset of tachyonic instability at the beginning is milder
(only a power divergence). For $p > 1$, the instability begins at $Z
< 0$ and it is much stronger (an exponential divergence). However,
in this region the nature of the tachyonic instability is
essentially the same for both $p\leq 1$ and $p>1$ cases. In other
words, the brane-antibrane system starting annihilation or tachyon
condensation takes place at a larger brane separation for $p \le 1$
case than for $p > 1$ case. Whether there is a deep reason or
implication behind this difference remains to be seen.\footnote{We
thank the referee for emphasizing to us the curious dependence on
$p$ for the onset of tachyonic instability in the $Z\to 0^+$ limit
which led to this discussion.}

\section{Evidence for the static nature of non-susy solutions}

 The static, non-supersymmetric and asymptotically flat $p$-brane
solutions\footnote{We use the terminology non-susy $p$-brane to
represent generically either the $p$-brane-anti$p$-brane system or
the non-BPS $p$-branes.} having isometries ISO($p,1$) $\times$
SO($d-p-1$) of type II supergravities in arbitrary space-time
dimensions ($d$) are given in \cite{Zhou:1999nm, Ivashchuk:2001ra,
Lu:2004ms}. For the purpose of this paper, we take $d = 10$ in the
following discussion. Unlike the BPS $p$-branes characterized by a
single unknown parameter, these solutions are characterized by three
unknown parameters and could be either charged or chargeless with
respect to a $(p+1)$-form gauge field. These non-susy $p$-branes
have a natural interpretation as coincident $p$-brane-anti-$p$-brane
(or non-BPS $p$-brane) \cite{Brax:2000cf,Lu:2004dp,Bai:2005jr,
Lu:2006rb}. As mentioned in the Introduction, this interpretation
has a puzzle since one would expect the non-existence of such static
solutions given the unstable nature of these systems.  We will use a
brane probe approach in this section to understand such static
nature of these configurations in the supergravity approximation.

The static non-supersymmetric $p$-brane solutions representing
coincident $p$-brane-anti $p$-brane systems in ten dimensional type
II supergravities are \cite{Lu:2004ms} \bea ds^2 &=&
F^{-\frac{7-p}{8}} \left(-dt^2 + dx_1^2 + \ldots + dx_p^2\right) +
F^{\frac{p+1}{8}}\left(H\tilde{H}\right)^{\frac{2}{7-p}} \left(dr^2
+ r^2 d\Omega_{8-p}^2\right)\nn e^{2\phi} &=&
F^{-a}\left(\frac{H}{\tilde{H}} \right)^{2\delta}\nn A_{[p+1]} &=& -
\sinh\theta \cosh\theta \left(\frac{C}{F}\right)dx^0 \wedge \ldots
\wedge dx^p \eea where we have expressed the metric in Einstein
frame. In the above, \bea F &=& \cosh^2\theta
\left(\frac{H}{\tilde{H}}\right)^\alpha - \sinh^2\theta
\left(\frac{\tilde{H}}{H}\right)^\beta\nn C &=&
\left(\frac{H}{{\tilde H}}\right)^\alpha - \left(\frac{{\tilde
H}}{H} \right)^\beta \nn H &=& 1 + \frac{\omega^{7-p}}{r^{7-p}},
\qquad \tilde{H}\,\,\,=\,\,\, 1 - \frac{\omega^{7-p}}{r^{7-p}} \eea
with the parameter relation \be b = (\alpha+\beta)(7 -p) \,g\,
\omega^{7 - p} \sinh2\theta \ee Here $\alpha$, $\beta$, $\theta$,
and $\omega$ are integration constants, $g$ is the string coupling
and $a=(p-3)/2$. Also $\alpha$ and $\beta$ can be solved, for the
consistency of the equations of motion,  in terms of $\delta$ as
\bea \alpha &=& \sqrt{\frac{2(8 - p)}{7 - p} - \frac{(7 - p)(p +
1)}{16} \delta^2} + \frac{a\delta}{2}\nn \beta &=& \sqrt{\frac{2(8 -
p)}{7 - p} - \frac{(7 - p)(p + 1)}{16} \delta^2} -
\frac{a\delta}{2}. \eea These two equations indicate that the
parameter $\delta$ is bounded as \be |\delta| \leq \frac{4}{7 - p}
\sqrt{\frac{2 (8 - p)}{p + 1}}.\ee The solution (14) is therefore
characterized by three parameters $\delta$, $\omega$ and $\theta$.

As demonstrated successfully in \cite{Lu:2004dp}, once the three
parameters of the above solutions are expressed in terms of the
number of D$p$-branes ($N$), the number of anti D$p$ branes ($\bar
N$) and tachyon parameter $T$, the tachyon condensation process can
be described correctly. In particular, we have the parameter
$\delta$ \be \delta = \frac{a}{|a|} \sqrt{\frac{8 - p}{2(7 -
p)}}\left[|a| \sqrt{\cos^2T + \frac{(N - \bar N)^2 }{4 N \bar N
\cos^2 T}} - \sqrt{a^2 \left(\cos^2 T + \frac{(N - \bar N)^2}{4
N\bar N \cos^2 T}\right) + 4 \sin^2 T}\right],\ee and the ADM mass
\be M (N, \bar N, T) = T_p \sqrt{(N+{\bar N})^2 - 4 N {\bar
N}(1-\cos^4 T)},\ee with $0 \le T \le \pi/2$ and $T_p$, the single
D$p$ brane tension. We have in general $M = (N + \bar N) T_p + V
(T)$ with $V (T)$ the tachyon potential. Just at the start of
tachyon condensation, i.e., $T = 0$, we should have $V ( T = 0 ) =
0$ and $M = (N + \bar N ) T_p$. The expression in (20) for $M$
indeed satisfies this. As discussed in \cite{Brax:2000cf}, the
parameter $\delta$ vanishes at $T = 0$ and the above expression in
(19) satisfies this, too. In order to examine whether the force
between the D$p$ and the $\bar {\rm D}p$ in the coincident
D$p$-$\bar {\rm D}p$ system vanishes or not using a probe D$p$ or a
probe $\bar{\rm D}p$, we need to place the probe at the location of
the coincident D$p$-$\bar {\rm D}p$ branes, i.e., at $r = \omega$ as
implied in (14). While at the same time, we need to make sure that
the probe remains as a probe, i.e., not strongly bounded to the
original coincident D$p$-$\bar {\rm D}p$ system. This can be so only
at the start of the tachyon condensation since there $M (N, \bar N,
T) = (N + \bar N) T_p$, or at least close to the top of the tachyon
potential where $M (N + 1, \bar N, T) \approx M (N, 1 + \bar N, T)
\approx T_p + M (N, \bar N, T)$. In other words, the value of the
parameter $\delta$ should be very close to its initial vanishing
value at the start of tachyon condensation. This will be the key
point for us to show in the following that the probe does indeed
feel `no force' when placed at the $r = \omega$.

For definiteness, let us consider a D$p$-brane probe placed at a
radial distance $r \ge \omega$ and parallel to the brane directions $x_\|$
in the static non-susy D$p$ brane background (14). Our purpose is to
calculate the interaction potential and for this we just need to
consider the bosonic worldvolume action for the probe and freeze the
worldvolume excitations as usually done for a $p$-brane probe placed
in a BPS $p$-brane background in \cite{Duff:1994an}. The bosonic
Lagrangian density for a D$p$ probe placed along $x_\|$ without
worldvolume excitations is \be {\cal L}_p = - T_p  \left[ e^{ -
\phi} \sqrt{- {\rm det} \,\gamma_{\mu\nu}} + A_{01\cdots p}\right]
\ee where\footnote{If the probe is anti D$p$ brane, we just need to
change the sign in front of $A_{01\cdots p}$ and the conclusion will
remain the same.} $\gamma_{\mu\nu} = g_{\mu\nu}$ with $g_{\mu\nu}$
the background metric (14) along $x_\|$ directions but now in the string
frame, and $A_{01\cdots p}$ is the corresponding Ramond-Ramond
background potential. Here we have set the worldvolume coordinates
$\sigma^\mu = X^\mu$ with $\mu = 0, 1, \cdots p$ and frozen the
worldvolume excitations. From the relation between string frame
metric and the one given in (14) which is in Einstein frame, we have
now \be g_{\mu\nu} = e^{\frac{\phi}{2}} \,F^{- \frac{7 - p}{8}}\,
\eta_{\mu\nu} = F^{- \frac{a}{4} - \frac{7 - p}{8}} \,
\left(\frac{H}{\tilde H}\right)^{\frac{\delta}{2}} \, \eta_{\mu\nu}=
F^{-\frac{1}{2}}\, \left(\frac{H}{\tilde
H}\right)^{\frac{\delta}{2}} \, \eta_{\mu\nu},\ee where we have used
the explicit background in (14) and $a = (p - 3)/2$. The potential
density (or the potential per $p$-brane volume) can be calculated
using background (14) as \bea V_p &=& T_p  \left[ e^{ - \phi}
\sqrt{- {\rm det} \,\gamma_{\mu\nu}} + A_{01\cdots p}\right]\nn &=&
T_p \left[F^{- 1} \left(\frac{H}{\tilde H}\right)^{\frac{a \delta
}{2}} - \sinh\theta \cosh\theta \left(\frac{C}{F}\right) \right]\nn
&=& T_p \frac{\left(\frac{\tilde H}{H}\right)^{\alpha -
\frac{a\delta}{2}} - \sinh\theta \cosh\theta (1 - \left(\frac{\tilde
H}{H}\right)^{\alpha + \beta})}{\cosh^2\theta - \left(\frac{\tilde
H}{H}\right)^{\alpha + \beta}\,\sinh^2\theta },\eea where we have
used the expressions for $F$ and $C$ in (15) in the last line above.
Note that from (17), $\alpha - a \delta/2 \ge 0$ and
$\alpha + \beta \ge 0$\footnote{$\alpha + \beta \ge 0$ guarantees $F
\ge 0$ and in that case the metric in (14) is well defined for $r \ge
\omega$ as noticed in \cite{Bai:2006vv}. The parameter $\delta = 0$
at $T = 0$ previously pointed out in \cite{Brax:2000cf} prompted our
discussion in \cite{Bai:2006vv} for the two disjoint decay channels
of brane-antibrane systems, one in terms of open string tachyon
condensation and the other in terms of the closed string tachyon
condensation. In the former case the system ends up
with a stable BPS configuration and while in the latter the system
ends up with "bubble of nothing" through black brane. We will discuss these
related issues in more detail elsewhere.}, therefore the potential
density remains finite for $r \ge \omega$, as can be seen from the explicit
expressions of $H$ and $\tilde H$  and their dependences on the
radial distance $r$ given in (15).

Given the above potential density, we can now calculate the force
per unit $p$-brane volume for the probe as \bea f_p &=& - \frac{d
V_p}{d r}\nn &=&  -   \frac{\alpha + \beta}{2}\, T_p
\frac{\cosh^2\theta + \left(\frac{\tilde H}{H}\right)^{\alpha +
\beta}\,\sinh^2\theta + 2 \left(\frac{\tilde H}{H}\right)^{(\alpha +
\beta)/2} \sinh\theta \cosh\theta}{\left[\cosh^2\theta -
\left(\frac{\tilde H}{H}\right)^{\alpha +
\beta}\,\sinh^2\theta\right]^2}\nn &\,& \times \left(\frac{\tilde
H}{H}\right)^{(\alpha + \beta)/2 - 1} \frac{2 (7 - p) \omega^{7 -
p}}{H^2 r^{8 - p}},\eea where we have used the explicit expressions
for $H$ and $\tilde H$ given in (15),  and also the relations for
$\alpha$ and $\beta$ in (17). Given $\alpha + \beta \ge 0$ and $ 0
\le \tilde H/H \le 1$ for $r \ge \omega$, it can be checked easily
that the above force is always attractive when $r > \omega$ as
expected while at $r = \omega$ could be either zero or divergent
depending solely on the sign of $(\alpha + \beta)/2 - 1$ since
$\tilde H/H = 0$ at $r = \omega$ (i.e. at the location of the
coincident D$p$-$\bar {\rm D}p$), and $1 \le H \le 2$ for $r \ge
\omega$. Now to show this we write from (17), \be \frac{\alpha +
\beta}{2} - 1 = \sqrt{\frac{2(8 - p)}{7 - p} - \frac{(7 - p)(p +
1)}{16} \delta^2} - 1.\ee From what we have discussed below (20)
regarding the validity of a probe when it is placed at $r = \omega$,
we know that the $\delta$ parameter should be very close to its
vanishing value at $T = 0$. With this and noting  $0 \le p \le 6$
for well-behaved supergravity solutions, one can check, for example,
taking $\delta = 0$ in the above that $(\alpha + \beta)/2 - 1 > 0$
for each allowed $p$. We thus notice from (24) that the force indeed
vanishes. This, therefore, gives evidence for the static nature of
the coincident D$p$-$\bar{\em D}p$ system in the supergravity
approximation as promised. We can actually do better for the
parameter $\delta$. Requiring $(\alpha + \beta)/2 - 1 > 0$, gives
\be |\delta| < \frac{4}{7 - p}\sqrt{\frac{9 - p}{p + 1}}.\ee This
bound allows the $\delta$ parameter to be deep in the bounded and
the tachyon condensation region (i.e., far away from its vanishing
value at $T = 0$ where the validity of the probe is guaranteed),
therefore, providing even further evidence for the static nature of
these solutions.

Given the property of a BPS $p$-brane supergravity configuration
that a (BPS) probe $p$-brane will feel no-force at any transverse
location when placed in this background and parallel to the large
number of coincident source $p$-branes \cite{Duff:1994an}, enables us
to obtain the stable BPS muti-$p$-brane configuration through a linear
superposition of individual BPS $p$-brane configuration at different
locations or placed coincidentally. With this, we expect that the
force acting on a probe $p$-brane placed parallel to the source branes
in the $p$-brane-anti $p$-brane background is due to the anti $p$-branes.
When the probe is placed at the same location as the coincident
branes in the $p$-brane-anti $p$-brane system, this probe brane acts the
same as the brane in the brane-anti brane system in
the tachyonic parameter region validating the probe approximation.
So if the
force acting on the probe vanishes, this may indicate that there is
no-force acting between the coincident branes and anti branes in
the brane-anti brane system in the supergravity approximation.
Therefore, this provides an evidence to support the existence of static
supergravity configuration describing the brane-anti brane system
in the supergravity approximation. This is the rationale behind what we
have shown above.

Now how to reconcile the result obtained here with the divergent force
calculated in the previous section\footnote{We again thank the
referee for raising a pertinent question which has led to
the discussion in this paragraph.}?
Before we address this, let us first point out the
differences between these two scenarios. The supergravity
description of the brane-antibrane system is obtained in the
supergravity approximation where we consider only the
massless modes in type II theories and their self-interaction
(back-reaction) in the lowest order approximation. While the force
calculation between a brane and an antibrane in the previous section
counts all the modes but no back-reaction and the divergence is due
to the tachyon mode when the brane separation is of the order of
string scale. It is not difficult to check that if we count only the
contribution of the massless modes to the force in the previous section (or
for large brane separation where only the massless modes
contribute), the result is always finite which is qualitatively
consistent with the above probe calculation. Even in this case, we
can only expect the two calculations to agree asymptotically in
which the backreaction can be ignored and if the two systems can be
prepared to be identical, i.e., a probe brane and a brane-antibrane
system with a given separation. But the force acting on the probe in
this section is evaluated at $r = \omega$, the location of the
coincident branes-antibranes, where the two calculations have no way
to agree, unlike the BPS case. So one should not directly compare
the force calculated in this section to the one in the previous section.

The probe approach used in this section serves only the purpose of
showing the static nature of brane-antibrane configuration in the
supergravity approximation and should not be taken as a
well-approximated calculation of brane-antibrane force in general.
The rationale for doing this is explained in the Introduction and we
will not repeat it here.

 In summary, we have used a probe approach to provide a direct
evidence to show that the force between the D$p$ and the ${\rm {\bar
D}}p$ in the coincident D$p$-$\bar{\rm D}p$ system in general
vanishes. This, therefore, justifies the static nature of the
general coincident D$p$-$\bar{\rm D}p$ configurations in
supergravity and such static nature is due to the supergravity
approximation.

\vspace{.5cm}

\noindent {\bf Acknowledgements}

\vspace{2pt}

The authors wish to thank the anonymous referee for the suggestions and
comments which has helped us, we hope, to improve the manuscript.
JXL acknowledges
support by grants from the Chinese Academy of Sciences and grants
from the NSF of China with Grant No:10588503 and 10535060.

\end{document}